# NOVEL BONDING TECHNOLOGIES FOR WAFER-LEVEL TRANSPARENT PACKAGING OF MOEMS

*Herwig Kirchberger*, *Paul Lindner, Markus Wimplinger*

EV Group, A-4782 St. Florian, DI Erich Thallner Str. 1, Tel +43 7712 5311-0, Fax +43 7712 5311 – 4600, E-Mail h.kirchberger@evgroup.com

**ABSTRACT**

Packaging costs of Micro-Electro-Mechanical System (MEMS) are still contributing with >50% to the total costs of most devices. Aligned wafer bonding techniques for Wafer-level packaging (WLP) demonstrates a huge potential to reduce these costs due to a smaller size of the total package, improved performance and shorter time to market. A special group of MEMS devices, Micro-Opto-Electro-Mechanical Systems (MOEMS), enjoys the potential to account for almost one third of the total MEMS sales by 2010, mainly driven by Digital Mirror Devices (DMD). MOEMS packages do require optical transparency at the operating wavelength of the individual sensors and actuators in addition to the conventional MEMS packaging requirements. This paper presents well understood wafer bonding and bond alignment technologies as well as high-volume proven equipment solutions for next generation transparent WLP of MOEMS.

## 1. INTRODUCTION

Each MEMS device category, its function and operational environment will individually dictate the packaging requirement. Innovation with regards to cost reduction and standardization in the field of packaging is therefore of utmost importance to the speed of commercialization of MEMS devices. The forecasted tremendous compound annual growth rate (CAGR) of MEMS device sales of around 13% over the next 5 years, which is mainly driven by devices for consumer applications, can drift upwards or downwards depending on the rate of innovation in the field of packaging.

MEMS devices typically require a specific fabrication process where the device wafer is bonded to a second wafer which effectively encapsulates the sensitive MEMS structure. This method leaves the device free to move within a vacuum or an inert gas atmosphere. These bonds are typically hermetic and therefore prevent moisture contamination and subsequent failure of the microstructure. In addition, MEMS devices can exhibit sensitivities that must be accounted for in the packaging solution. Device performance can be affected by stresses imparted to the body of the die by the packaging materials. Emerging MEMS devices are going to require device-specific packaging far into the foreseeable future. Of course protection is a key element in packaging of MEMS because corrosion, moisture, and debris can prevent the devices from working.

Wafer Bonding for Wafer-level packaging (WLP) is a technology where the entire packaging process is performed at wafer level. Although the basic WLP infrastructure has been available for several years the technology was not adopted since existing packaging solutions met market requirements. WLP can provide a solution when requirements for continued reduction in size and required cost targets are not met by traditional packaging. Today most MEMS companies have already transitioned away from die level packaging.

Reducing the package costs is the most efficient approach to reduce the total device costs for many consumer electronics devices. Disruptive manufacturing innovations like wafer-level-packaging instead of chip-level packaging therefore give a huge competitive advantage.

## 2. MICRO-OPTO-ELECTRO MECHANICAL SYSTEMS (MOEMS)

Turning the focus towards the fastest growing MEMS market segment, MOEMS with its leading representative DMD, it is well understood that innovative packaging technologies have a major impact on the final device costs and therefore on the commercialization of the final product. The global MOEMS market, mainly driven by DMD sales, is expected to overtake the inkjet head market as the largest MEMS device segment by the end of 2006 and by 2010 it is forecasted that MOEMS will account for almost one third of the total MEMS device sales.
With regards to wafer-level packaging, MOEMS packages do require optical transparency at the operating wavelength of the individual sensors and actuators in addition to the conventional MEMS packaging requirements including electrical interconnectivity, hermetic enclosure, protection from micro-contamination, mechanical stress reduction and the elimination of stiction.

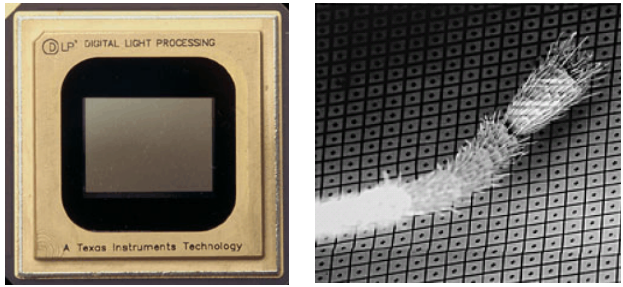

**Fig.1.** DLP¨ Chip (Digital Light Processing Chip) for XGA from Texas Instruments (left); leg of an ant on a DMD surface (right)

The DLP¨ chip is used in display systems, home entertainment systems and high resolution imaging systems and is containing far above 1 Million tiny moving mirrors. Each mirror is addressable and reflects light in and out of a focusing assembly. Each mirror is sitting on top of a torsional spring and is actuated electrostatically, tilting each mirror between 2 defined states (-10°/+10°) within less than 16ms. This fast switching allows grays to be emulated, 3 DMD arrays respectively a wheel is used to create color. MEMS mirrors are built upon a defined CMOS circuitry using surface micromachining of several aluminum layers. Most critical are the deposition of a low stress torsion hinge and the placement of the landing pad. Sacrificial material is organic and is removed by an oxide plasma etching process. Major failure mechanisms are surface contamination and "hinge memoryÓcaused by metal creep (refer also to http://www.dlp.com/).

These considerations on DMD functionality already demonstrate the sensitiveness and complexity that needs to be addressed in the required package of MOEMS devices.

### 3. WAFER BONDING TECHNOLOGIES

Wafer bonding is currently used for zero- and first-level packaging and has to compete on the cost level with established and also new backend techniques. There are several important points for the successful implementation of wafer bonding for wafer-level packaging.

*Process transfer from R&D to production*
Due to the very short product life cycles in consumer electronics, sometimes only a few months, a smooth transfer from research and process development into production is extremely important.

*Capacity enhancement*
In many cases a specific packaging process can be transferred from one product to other similar devices. The wafer bonding platform should be designed in such a way that a capacity enhancement can be easily accommodated. In consumer electronics the lifetime of semiconductor equipment outlasts the device product life times by magnitudes. A universal bond chamber gives full flexibility to switch between different wafer bonding processes without additional capital equipment expenses.

*Processes independence*
In order to minimize the costs for process development and qualification it is desirable to standardize the individual process steps. Novel wafer-to-wafer alignment techniques enable standardization independent of wafer material, size and properties.

*Yield enhancement*
Most wafer bonding applications are very sensitive with regards to the wafer surface parameters. In order to achieve a high yield manufacturing process it is therefore necessary to control the surface properties as tightly as possible. The key pre-processes for wafer bonding are cleaning, plasma activation and adhesive layer coating. Integration of the pre-processes in the wafer bonding platform enables real-time process control. Due to the accurate timing of the process flow the bonding process time can be optimized and the capacity and throughput are increased.

The most common permanent wafer bonding technologies are subsequently reviewed and evaluated for its suitability for transparent WLP for MOEMS.

*Anodic Bonding*
It is the most widely used packaging method up to now. Electric fields assist in the thermal diffusion of ions across the bond interface to achieve solid state mixing of the optical transparent glass and the silicon. Anodic bonds are very common in MOEMS because of the transparent glass lid. Anodic bonding is considered the workhorse of MEMS packaging and accounts for the majority of all packaging applications. Advancements in wafer clamping and independent thermal processes for top and bottom chucks are key to managing thermal expansion differences and maintaining submicron alignment during whole wafer processing.

*Silicon Direct Bonding*
Numerous materials can be bonded using this technique provided the surfaces meet the rigid roughness and flatness standards. Hydrophilic surfaces created by wet chemistry or plasma activation will immediately bond upon contact via Van-der-Waals attractions between adsorbed water groups. Following the bond, batch thermal annealing is used to transition the bonds to covalent Si-O-Si bonds and achieve bond strengths equivalent to bulk silicon. Because silicon direct bonding requires atomic intimacy at the interface, point-of-use cleaning methods are essential in high volume production. Modern cluster tools (refer to the EVG GEMINI© system in Fig.3) capable of cleaning, activation, aligning and bonding can achieve 25-40 wafer/hour throughputs and are contained in class 1 mini-environments.

Silicon Direct Bonding is a method that allows for photosensitive elements to be illuminated from the back side, thus achieving higher efficiencies through larger optical active areas. After bonding of the product wafer face down to a second Si wafer, the material can be backgrinded to Si thicknesses which are sufficiently optical transparent.

*Thermal Compression Bonding*
This includes three main subcategories; glass frit, eutectic, and diffusion.

*Glass Frit Bonding*
During glass frit bonding the intermediate layer at the interface begins to flow under the influence of pressure when heated above the glass softening temperature. The glasses can be applied via extrusion, screen-printing, spraying, or sedimentation methods.

*Eutectic Bonding*
Eutectic bonding takes advantages of a metallurgical phase transformation in which the binary phase formed from constituents has a lower melting point than either solute. Bonding techniques using metals as intermediate layers typically form a hermetic seal and are high vacuum compatible (low out gassing materials with low permeabilities).

In essence, eutectic bonding is a special case of a diffusion bond that allows for very strong inter-metallic bonds to be formed at relatively low temperatures. When two materials diffuse a mixture is formed. This mixture has a very low melting point at the eutectic composition. Once the eutectic forms and becomes liquid the reaction accelerates under the influence of liquid phase diffusion at the liquid solid interface.

Solid-state thermo compression bonding is similar to eutectic bonding because an alloy is formed. However, these reactions do not involve melting of the diffused interface layer. In solid state bonding, the key is to identify systems with low temperature solid-state phase transformations and rapid diffusion coefficients. The phase that forms is most often an inter-metallic compound that is very tough and gives structural stability to the assembly.

*Diffusion Bonding*
This type of thermal compression bonding is generally applicable to systems in which the diffusion coefficient is rapid at relatively low temperature. This occurs for some metals such as Au and Cu. Thus it is possible to create Au-Au and Cu-Cu bonds or even Cu-Au bonds at low temperature using diffusion driven kinetics. In these cases, no alloys are formed and the interface is a mixture of the two solutes. In some applications, a diffusion bond is preferable to inter-metallic or eutectic alloy formation because of the generally brittle nature of those alloys.

In general Thermal Compression bonding is also suitable for transparent MOEMS packaging assuming sufficient temperature budget and stability of the package (most of the glasses are temperature stable up to 500°C and can also be metallized) and assuming that the intermediate layers (glass frit, metals) can be arranged in frames around the active area of the MOEMS device (e.g. screen printing or masked CVD). Nevertheless this type of bonding is not very common in MOEMS packaging so far since it exhibits higher costs than comparable bonding techniques. Nevertheless if ultra-low leakage rates need to be considered, thermal compression bonding is the best alternative.

*Adhesive bonding*
Adhesive bonding with UV curable and transparent coatings and low-k dielectrics like Benzo-Cyclo-Butene (BCB) represent the majority of packaging techniques for MOEMS. Ease of application, low material costs and sufficient bond strength as well as permeability contribute to the popularity of adhesive bonding for the transparent packaging of MOEMS.

Low-k dielectric adhesives gain more and more attention also for the packaging of MOEMS since they allow for the creation of electrical interconnects between different functional modules, which is a necessity for a true WLP, and can be patterned by means of Lithography. This allows for the cheap formation of bonding frames around the active area of the MOEMS.

## 3. WAFER LEVEL TRANSPARENT PACKAGING METHODS FOR MOEMS

Wafer Level Bonding utilizing optical grade materials found its breakthrough with packaging of CMOS image sensors in the 1990s. Shellcase was first in licensing chip-scale wafer-level packaging technologies (ShellOP<sup>TM</sup>, ShellOC<sup>TM</sup>, ShellUT<sup>TM</sup>) to several packaging companies in Asia. Other major players followed soon, developing their own transparent packaging technology and enjoying a strong increase in sales in the fast growing consumer markets (especially camera chips for mobile phones).

Meanwhile a number of key players in this field are starting to implement additional functionality into the transparent package. Multiple Glass wafers, stacked on top of each other with integrated micro-lens structures to realize a fully integrated camera package seem to be the Holy Grail in this industry nowadays. The achievable precision with regards to the structural integrity of each individual micro-lens, the overall Alignment Accuracy of the transparent wafer stack and the introduced stress in the glass wafer stack during low temperature wafer bonding still need to be improved to implement the first fully integrated optical camera package into mobile phones.

All those applications require a low temperature wafer bonding method with high alignment accuracy. This is usually accomplished by either using a thermal or UV curable Epoxy as intermediate layer and a separate optical alignment step before bonding the clamped wafer pair under vacuum or under a controlled atmosphere, temperature and pressure. The crucial process step is to maintain the pre-bond alignment accuracy after bonding, since the Epoxy intermediate layer exhibits a certain flow characteristic before curing. The reliable clamping of the optically aligned wafers, which are separated by spacers, is the key to success. The transfer process from the bond aligner to the wafer bonding system needs to be controlled very accurately. Another challenge, which is addressed by special tooling solutions, is the lack of flatness of the transparent cap wafer. The equipment needs to flatten the wafer sufficiently in order to maintain the alignment accuracy after the curing process of the intermediate layer.

For Si and several other materials, low temperature dry activation (LTDA) of the wafers prior to bonding improves the bonding strength at temperatures well below 300°C drastically (Fig.2). Yet another solution is to use high performance transparent polymer wafers. In this case the bond can be achieved by either using curable intermediate layers or direct polymer bonding methods assisted by low temperature plasma activation. LTDA has proven to allow direct polymer wafer bonding close to room temperature.

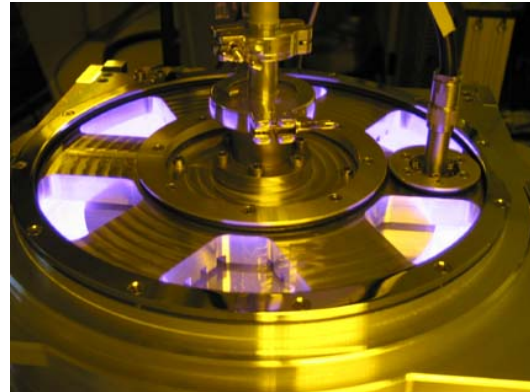

**Fig.2.** Chamber detail of an EVG810 Plasma Activation tool. The closed chamber design allows for highest plasma uniformities

For MOEMS with moving sensors and actuators special attention needs to be paid to the protection and functional integrity of these parts.
Fully automated cluster tools, which are integrating bond alignment and wafer bonding modules as well as modules for intermediate layer curing with UV or surface activation for direct bonding methods (Fig.3) can be equipped with inert gas ECUs (Environmental Control Units) to maintain a dry atmosphere during encapsulation. In order to prevent a contact of the intermediate layer with the active device areas usually bonding frames around these areas are formed with screenprinting or micro-contact printing. Novel cluster tools like the GEMINI¨ system can implement such pre-process modules as well. Another very common-way to prevent contamination of the active area, especially for fragile and movable MOEMS devices, is the use of an interposer wafer. Basically it is a third wafer which is incorporated in the package between the MOEMS wafer and the lid and which function can be described as wafer level array of bonding frames around the active areas. Usually this wafer is bonded first to the lid wafer and the resulting stack is then optically aligned and bonded to the MOEMS wafer. When using Silicon interposer and a glass lid the process can be accomplished by a simple and cheap mechanically aligned anodic bond.

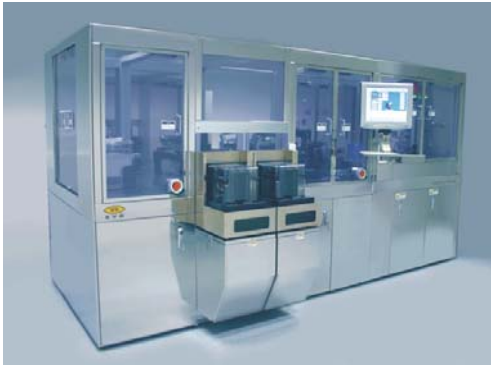

**Fig.3.** Fully automated production wafer bonder EVG Gemini¨ with SmartView¨ alignment and LowTemp¨ plasma activation module

Post Bond Alignment Accuracies with UV curable intermediate layers are typically in the range of 10µm@3Sigma due to the flow characteristics of the intermediate adhesive before curing. For direct bonding methods with plasma assisted fusion bonding, post bond alignment accuracy <1µm@3Sigma was successfully shown with SmartView¨ Alignment.

SmartView¨ alignment (Fig.4) employs two microscopes for alignment instead of using a single microscope in between the wafers. One microscope is placed above and the other below the wafer stack. The dual microscopes focus on a common focal plane calibrated for each alignment. Each microscope objective observes one alignment key on the surface of the wafer.

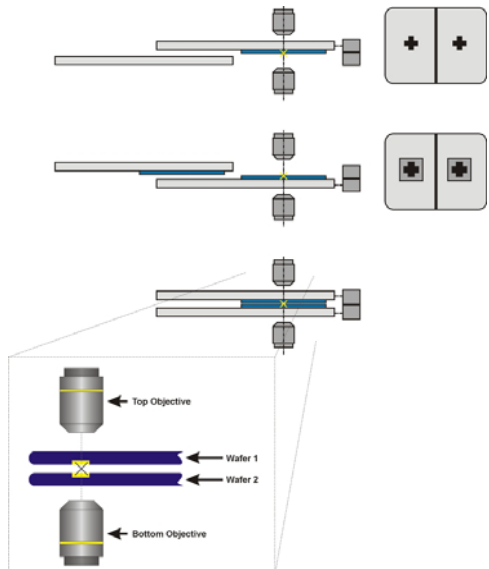

**Fig.4:** SmartView¨ alignment principle

Wafer alignment is accomplished using encoded stage motors allowing X and Y movements in increments of 0.1 µm steps and minimized Z-axis travel controlled by three software controlled spindle motors to preserve planarization between the top and bottom wafers.

### 3. CONCLUSIONS

Disruptive manufacturing innovations like wafer-level-packaging instead of chip-level packaging present a huge cost saving advantage for MEMS devices and allow their vast adoption into the consumer electronics market. MOEMS packages do require optical transparency at the operating wavelength of the individual sensors and actuators in addition to the conventional MEMS packaging requirements including electrical interconnectivity, hermetic enclosure, protection from micro-contamination, mechanical stress reduction and the elimination of stiction. For MOEMS with moving sensors and actuators special attention needs to be paid to the protection and functional integrity of these parts.
Fully automated cluster tools, which are integrating bond alignment and wafer bonding modules as well as modules for intermediate layer curing with UV or surface activation for direct bonding methods can be equipped with inert gas ECUs (Environmental Control Units) to maintain a dry atmosphere during encapsulation. Such cluster tools achieve 25-40 wafer/hour throughputs and the resulting alignment accuracy of the package can be in the range of <1µm@3Sigma, depending on the type of bonding.

### 5. APPENDIX

The table in the appendix illustrates the main process parameters for different wafer bonding processes used in MEMS wafer-level packaging.

| General method | | Electric Field | Direct bonding | | | Intermediate layer | | | | |
|---|---|---|---|---|---|---|---|---|---|---|
| Bond process | | Anodic | SDB | Plasma activated | Glass frit | Thermo-compression (metal-metal) | Eutectic | Adhesive | Epoxy (thermally cured) | Epoxy (UV cured) |
| CMOS compatible | | ✖ | ✖ | ✓ | ✖ | ✓ | ☑ | ☑ | ✓ | ✓ |
| Temperature range | ≤1000°C | ✖ | ✓ | ✖ | ✖ | ✖ | ✓ | ✖ | ✖ | ✖ |
| | <440°C | ✓ | ✖ | ✓ | ✓ | ✓ | ✓ | ✓ | ✓ | ✓ |
| | <200°C | ✖ | ✖ | ✓ | ✖ | ☑ | ✓ | ✓ | ✓ | ✓ |
| | RT | ✖ | ✖ | ☑ | ✖ | ✖ | ✖ | ✖ | ✖ | ✓ |
| Vacuum compatibility | Low <1mbar | ✓ | ✓ | ✓ | ✓ | ✓ | ✓ | ✓ | ✓ | ✖ |
| | High <10⁻²mbar | ✓ | ✓ | ☑ | ✖ | ✓ | ✓ | ✖ | ✖ | ✖ |
| Surface roughness (both surfaces) | <1µm | ✖ | ✖ | ✖ | ✓ | ✖ | ✓ | ✓ | ✓ | ✓ |
| | <20nm | ✖ | ✖ | ✖ | ✓ | ✖ | ✓ | ✓ | ✓ | ✓ |
| | <2nm | ✓ | ✖ | ✓ | ✓ | ✓ | ✓ | ✓ | ✓ | ✓ |
| | <0.5nm | ✓ | ✓ | ✓ | ✓ | ✓ | ✓ | ✓ | ✓ | ✓ |
| Environment | Cleanroom class | 100 | 10 | 10 | 1000 | 10 | 100 | 100 | 1000 | 1000 |
| Sensitivity to particles | Low | ✖ | ✖ | ✖ | ✓ | ✖ | ✖ | ✓ | ✓ | ✓ |
| | Medium | ✓ | ✖ | ✖ | ✖ | ✓ | ✓ | ✓ | ✖ | ✖ |
| | High | ✖ | ✓ | ✓ | ✖ | ✖ | ✖ | ✖ | ✖ | ✖ |
| Industrial environment | Low Volume | ✓ | ✓ | R&D | ✓ | ✖ | ✖ | ✓ | ✓ | ✓ |
| | High Volume | ✓ | ✓ | ✖ | ✓ | ✓ | ✓ | ✓ | ✓ | ✖ |

✓ : fully compliant
☑ : compliant with certain limitations or boundary conditions
✖ : not compliant